\documentclass[english,10pt, conference]{IEEEtran}
 \usepackage[pdftex]{graphicx}
 \usepackage[cmex10]{amsmath}
 \usepackage{booktabs}
\usepackage{multicol}
\usepackage{multirow}
\usepackage{caption}
\usepackage{array}
\usepackage[utf8]{inputenc}
\usepackage[T1]{fontenc}
\usepackage{microtype}
\usepackage{tablefootnote}
\usepackage{url}
\usepackage{amssymb}
\usepackage{latexsym}
\usepackage{verbatim}
\usepackage{array}
\usepackage{float}
\usepackage{booktabs}
\usepackage{adjustbox}
\usepackage[all]{nowidow}

\newcommand{\ie}[0]{{{i.e.,}\ }}
\newcommand{\eg}[0]{{{e.g.,}\  }}
\newcommand{\etal}[0]{{{et al.}\ }}
\newcommand{\cc}[0]{\cellcolor{white}}

\usepackage{color, colortbl}
\usepackage[table]{xcolor}
\definecolor{gray}{gray}{0.9}

\usepackage{cite}

\ifCLASSINFOpdf
\else
\fi
\usepackage[tight,footnotesize]{subfigure}

\begin{document}
%
\title{Rationale in Development Chat Messages:\\ An Exploratory Study}




%

\author{\IEEEauthorblockN{Rana Alkadhi\IEEEauthorrefmark{1},
Teodora La\c{t}a\IEEEauthorrefmark{1},
Emitza Guzman\IEEEauthorrefmark{2} and
Bernd Bruegge\IEEEauthorrefmark{1}}\\
\IEEEauthorblockA{\IEEEauthorrefmark{1}Faculty of Informatics,
Technische Universit{\"a}t M{\"u}nchen\\
Garching, Germany\\ \{alkadhi, teodora, bruegge\}@in.tum.de}

\IEEEauthorblockA{\IEEEauthorrefmark{2}Department of Informatics,
University of Zurich\\
Switzerland\\
guzman@ifi.uzh.ch}
}

\maketitle

\begin{abstract}

Chat messages of development teams play an increasingly significant role in software development, having replaced emails in some cases. Chat messages contain information about discussed issues, considered alternatives and argumentation leading to the decisions made during software development. These elements, defined as \textit{rationale}, are invaluable during software evolution for documenting and reusing development knowledge. Rationale is also essential for coping with changes and for effective maintenance of the software system. However, exploiting the rationale hidden in the chat messages is challenging due to the high volume of unstructured messages covering a wide range of topics. 
This work presents the results of an exploratory study examining the frequency of rationale in chat messages, the completeness of the available rationale and the potential of automatic techniques for rationale extraction. For this purpose, we apply content analysis and machine learning techniques on more than 8,700 chat messages from three software development projects. Our results show that chat messages are a rich source of rationale and that machine learning is a promising technique for detecting rationale and identifying different rationale elements.

\end{abstract}

\section{Introduction}  
\label{sec:intro}   
\begin{table*}
\centering
\captionof{table}{Definitions of rationale elements (adapted from~\cite{Bruegge2009}).} 
\begin{tabular} { p{3cm}   p{12cm}  }
\toprule
   \textbf{Rationale element}        & \textbf{Definition} \\ \midrule
   \rowcolor{gray}
   Issue  & Problem that needs discussion and negotiation to be solved. An issue typically can not be resolved algorithmically and does not have a single correct solution.\\
   
   Alternative & Possible solution that could address the issue under consideration.\\
   \rowcolor{gray}
   Pro-argument & Positive reason supporting an alternative.\\
   
   Con-argument & Negative reason against an alternative.\\
   \rowcolor{gray}
   Decision & The alternative selected to resolve an open issue.\\
   
   \bottomrule               

\end{tabular}
\label{tab:rationale_elements}
\end{table*}

Development teams make various decisions throughout the software lifecycle. They discuss current issues, propose alternative solutions, argue for and against these alternatives and collaboratively make decisions. All  elements justifying the made decisions constitute \textit{rationale}. 
The presence of rationale is invaluable during software evolution. For example, the availability of rationale improves the traceability, documentation and understandability of the system~\cite{Dutoit2006}. Moreover, documented and accessible rationale is essential for effective maintenance and for analyzing the impact of changes~\cite{Burge2008,Bruegge2009}. However, different methods for explicit rationale capture, \eg involving designers and developers in writing up their rationale in a formal schema, have been met with resistance. Possible explanations for their reluctance are the intrusiveness of these methods and time constraints~\cite{Dutoit2006, Fischer1991}.  Nevertheless, although rationale is rarely captured in a structured, explicit form, it is embedded in different development artifacts and communication channels~\cite{Rogers2014, Brunet2014}. To capture rationale from these artifacts, sources of rationale need to be identified and techniques for extracting rationale need to be developed~\cite{Rogers2012}.

One potential medium for rationale are chat messages exchanged between members of development teams. Chat messages are an integral part of the daily activities in software development ~\cite{Lin2016}. 
Team members discuss a wide range of topics in their chat messages including social, managerial, organizational and development issues. Some of this information could be an invaluable source of rationale. However, to the best of our knowledge, the extent of rationale in chat messages from development teams has not yet been explored. This work is a first effort in this direction. 

  
We report on an exploratory study to better understand the rationale present in chat messages, in regards to its \textit{frequency}, \textit{completeness} and the potential of machine learning techniques for \textit{automatically extracting rationale} elements on two granularity levels.  For the purpose of our study we manually analyzed 8,702 chat messages of three development teams using content analysis techniques~\cite{Neuendorf2002} and used the manually labeled data to train and evaluate different machine learning techniques.

Our results provide quantitative evidence that  chat messages from development teams are a valuable source for rationale. However, due to the high volume of chat messages and sparsity of the messages containing rationale, the manual extraction and classification of rationale is a tedious and time-consuming process. Our results show that machine learning techniques are promising for detecting rationale in chat messages with a recall ranging from 0.61 up to 0.88 and for filtering messages without rationale with a recall of up to 0.98.

The contribution of the study is threefold. First, we investigated the frequency and completeness of  rationale present in chat messages by using content analysis techniques and descriptive statistics. Second, we studied the performance of machine learning techniques for detecting rationale on the manually analyzed data. Third, we analyzed the potential of machine learning techniques for classifying rationale into different elements, \eg issues, alternatives and decisions. 

These contributions shed light on the rationale knowledge contained in software developers' chat messages and provide insights that could aid future research in exploiting this tacit knowledge.

\section{Rationale Representation Models}  
\label{sec:background}

\textit{Rationale} captures the reasons behind decisions. It includes the justification behind decisions, other alternatives considered and the argumentation that led to the decision~\cite{Lee1997}. 

Since Kunz and Rittel~\cite{KunzundRittel1970} proposed to capture rationale as an issue model, many models have been proposed, such as IBIS (Issue Based Information System)~\cite{KunzundRittel1970}, QOC (Question, Option and Criteria)~\cite{MacLean1991}, PHI (Procedural Hierarchy of Issues)~\cite{mccall1987} and DRL (Decision Representation Language)~\cite{Lee1990}. The rationale elements we analyze throughout this paper are based on IBIS for its conciseness and as it provides the basis for most of the subsequent issue models including DRL and QOC. Namely, we focus our  analysis on five elements: issues, alternatives, pro-arguments, con-arguments and decisions. Table~\ref{tab:rationale_elements} lists the rationale elements used in this work and their definitions. The rationale elements definitions were adapted from Bruegge and Dutoit~\cite{Bruegge2009}. These rationale elements are basic elements shared between most of the rationale representation models.



\section{Research Setting}  
\label{sec:research_setting}

In this section, we introduce our research questions, describe the followed research method and the data we used to perform our analysis. 

\subsection{Research Questions}



The aim of our study is to evaluate chat messages of development teams as a potential source for rationale. For this purpose,  we explored the rationale frequency in chat messages, the rationale completeness and the rationale automatic extraction potential from chat messages.



\textbf{RQ1: Rationale frequency} describes how often rationale appears in chat messages. This information gives a first insight of the worth of considering chat messages as a source of rationale. In particular, we answer the following question:
\begin{itemize}
    \item What is the frequency of rationale in chat messages?
\end{itemize}

\textbf{RQ2: Rationale completeness} describes the quality of the recovered rationale from chat messages. Rationale is distributed across different development artifacts, \eg bug reports~\cite{Rogers2014} and design session transcripts~\cite{Hesseb2016}, and all of these sources could be used together to capture a more complete rationale of the software system. 
\textit{Rationale completeness} occurs when for each documented decision, all the rationale elements justifying the decision are documented~\cite{BurgeJ2008}.  Answering this question could help practitioners and researchers by providing insights on how to integrate the rationale extracted from chat messages with the rationale extracted from other sources. In particular, we answer the following question:
\begin{itemize}
  \item How complete are the  rationale elements extracted from chat messages?
\end{itemize}



 
 
 \textbf{RQ3: Rationale automatic extraction} describes the potential of applying automatic techniques to detect and extract rationale from chat messages. This information provides insights on the feasibility of applying techniques previously used for retrieving important information  from other development artifacts to extract rationale from chat messages. 
 In particular, we answer the following questions:
 \begin{itemize}
    \item \textit{Binary classification:} Can rationale be accurately detected in chat messages by applying supervised machine learning techniques?
    \item \textit{Fine-grained classification:} Can chat messages containing rationale be accurately classified into the different rationale elements: issues,  alternatives,  pro-arguments,  con-arguments and decisions by applying supervised machine learning techniques?
    
\end{itemize}
 
 
 \begin{table}
\centering
\captionof{table}{Overview of chat messages dataset. } 
\begin{adjustbox}{max width=0.45\textwidth}
\begin{tabular}{ lrr}
\toprule
   \textbf{Team}    & \textbf{Chat messages before filtering}         & \textbf{Chat messages after filtering}   \\ \midrule
   \rowcolor{gray} 
 \textbf{Team A} & 4,106 & 3,974\\
 \textbf{Team B} &  2,214 & 2,164\\
 \rowcolor{gray} 
 \textbf{Team C} & 3,026 & 2,564\\
 \hline
  \textbf{Total}& \textbf{9,346} & \textbf{8,702} \\
\hline

\end{tabular}
\end{adjustbox}
\label{tab:dataset}
\end{table}
 
 \subsection{Research Method}
 We employed content analysis techniques~\cite{Neuendorf2002} to study the \textbf{rationale frequency} in chat messages and the \textbf{rationale completeness} of the existing rationale. Our analysis consisted of two consecutive steps. First, we identified the rationale elements present in the chat messages. Second, we evaluated the completeness of the identified rationale by linking semantically related rationale elements. Finally, to study the \textbf{rationale automation extraction} we applied machine learning techniques on the manually analyzed data and evaluated the classification performance of these techniques according to well established metrics. We explain each procedure in further detail in Sections~\ref{sec:frequency}, \ref{sec:completeness}, \ref{sec:automatic_extraction}.

\subsection{Research Data}

We analyzed the chat messages of three development teams that were part of a multi-project capstone course at the Technical University of Munich in 2015 and 2016~\cite{Bruegge2012, Krusche2014}. The course was designed to simulate industry settings~\cite{Krusche2014b}. During the course, teams developed mobile applications for industrial partners and dealt with incomplete and evolving requirements following an agile software methodology~\cite{Krusche2014b}. The participants used Atlassian HipChat for instant messaging\footnote{https://www.hipchat.com}, which supports diverse integrations to external services and bots, \eg with Bamboo\footnote{https://www.atlassian.com/software/bamboo} to send notifications about build results and Standup Bot\footnote{http://botlab.hipch.at} to report and retrieve statuses for stand-up. Each team consisted of 8 to 9 students and a project leader for project management. When selecting the teams for the study, we considered only teams who communicated in English and wrote more than 2,000 messages.  To focus our analysis on the messages written by members of development teams, messages that were automatically generated by one of the services or bots were filtered out. Table~\ref{tab:dataset} shows the three selected teams and the number of chat messages before and after filtering automatically generated messages. 

\section{Rationale Frequency}
\label{sec:frequency}


To explore the  frequency  of  rationale  in  chat messages, we analyzed how often rationale appears in chat messages and the frequency of  different rationale elements.


\subsection{Procedure} 
\begin{figure} 
 	 \includegraphics[width=0.45\textwidth]{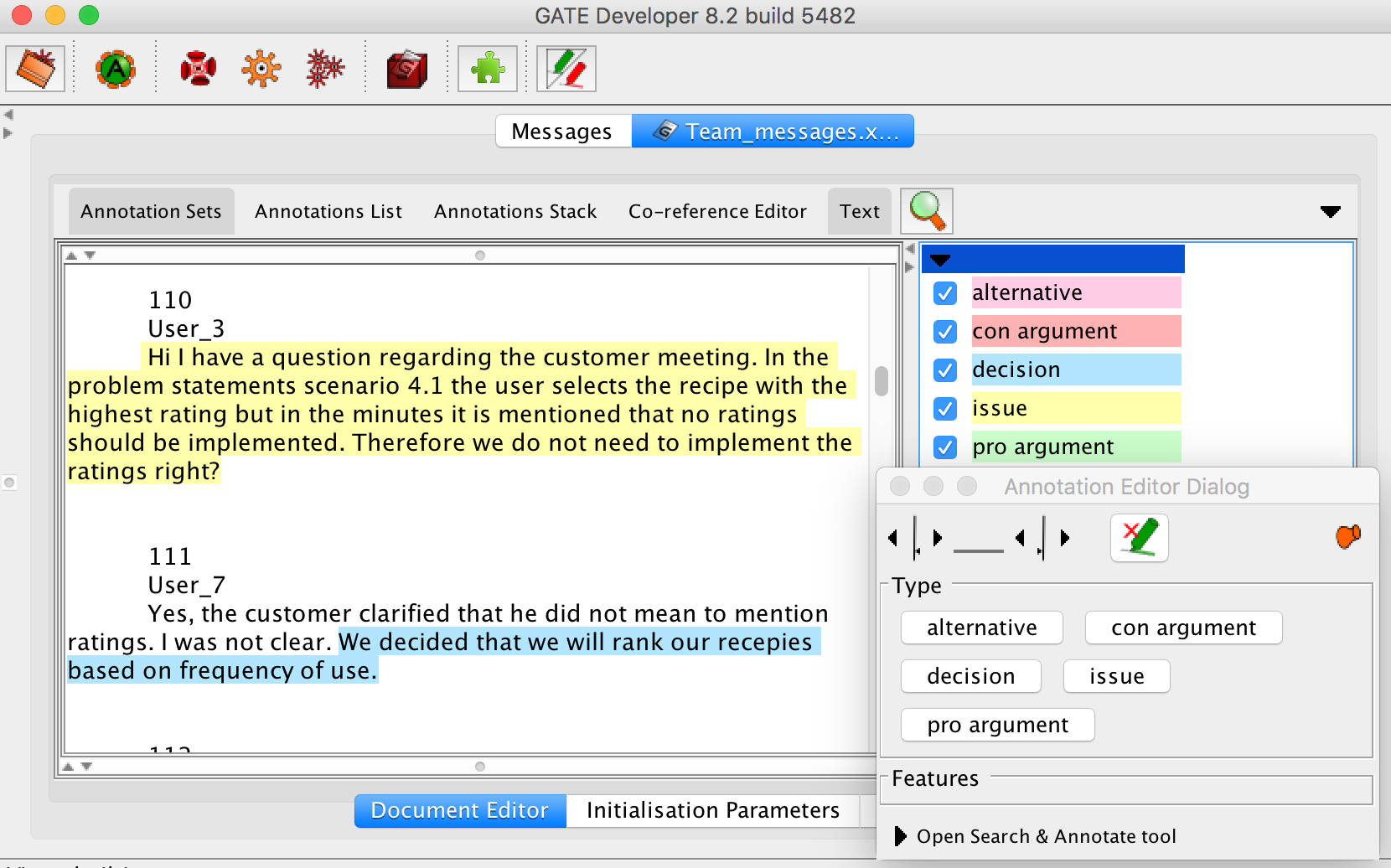}
 	 \caption{A screenshot of using GATE for the manual coding of chat messages.}
 	 \label{fig:GATE}
 \end{figure}  
We manually analyzed the chat messages in our dataset by applying content analysis techniques~\cite{Neuendorf2002}. Two of the authors of this paper independently inspected the chat messages of the three teams in our dataset and identified the contained rationale. This process comprised three steps:
\subsubsection{Developing the coding guide} The aim of this step was to systematize and minimize disagreements between the two coders. Since many representations have been proposed in literature to model rationale, it is important that the two coders share a unified understanding of the elements that constitute rationale. To this end, we developed a coding guide that includes clear definitions of the rationale elements and examples for each element\footnote{Available on: https://goo.gl/PAKLQU}. 
Table~\ref{tab:rationale_elements} lists the rationale elements used in this paper and their definitions. The coding guide was developed in an iterative process consisting of two trial iterations. In each iteration, the two coders independently identified rationale elements discussed in 1,000 chat messages.  The coders disagreements were analyzed and the coding guide was modified to minimize disagreements in the next iterations. 

\subsubsection{Coding of chat messages} For the manual coding task, we used GATE (General Architecture for Text Engineering)~\cite{Cunningham2011a,Cunningham2013}, a Java-based framework for a diverse set of natural language processing applications. 

Figure~\ref{fig:GATE} shows a screenshot of GATE as used by coders. The main window displays the list of chat messages to be coded. If a message contains rationale, the coder highlights the message part containing the rationale and specifies its type.

The coding unit, \ie the highlighted part, can be one sentence, multiple sentences of a message or the complete message. We refer to the coded units as \textit{text snippets}.  For each snippet, coders specified whether it contains rationale and what type of rationale elements are present. A text snippet might contain multiple rationale elements. 
Coding text snippets allows for capturing the text containing rationale in the finest-grained manner since a message might contain additional irrelevant information.




The two coders independently coded each message in our dataset. The average time to code 8,702 messages was 13 hours per coder, highlighting the large effort required to manually extract rationale elements. 
 
\subsubsection{Disagreement reconciliation}
Disagreements between the two coders included situations when the two coders identified different rationale elements in the same text-snippet or when only one coder coded a text-snippet as containing rationale. The disagreements where resolved through discussions between the two coders.
\begin{figure} 
 	 \includegraphics[width=0.45\textwidth]{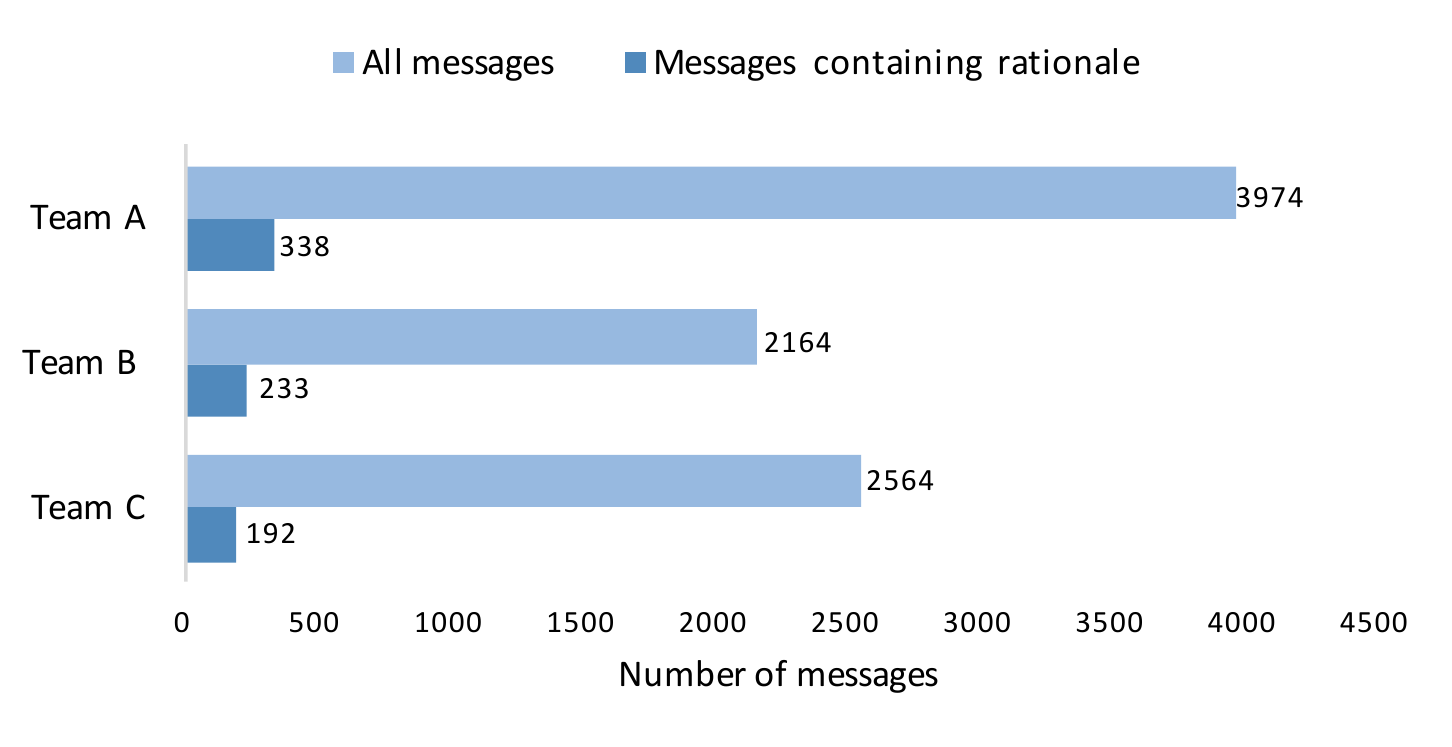}
 	 \caption{Chat messages containing rationale per team.}
 	 \label{fig:relevant_msgs}
 \end{figure} 

\begin{table*}
\centering
\captionof{table}{Frequency of rationale elements across messages containing rationale per team. } 
\def\arraystretch{1.5} 
\begin{tabular}{ lrrrr p{7 cm} }
\toprule
& \multicolumn{4}{c}{\textbf{Frequency}} & \\
\textbf{Rationale element} & \textbf{Team A} & \textbf{Team B} & \textbf{Team C} & \textbf{Total} & \textbf{Example} \\
\midrule
Issue &25\% & 28\% & 17\% &24\%& \textit{``Plus if this is implemented using segueways, what screen do you go back to when you click 'back'?''}\\
Alternative & 45\% & 54\% & 57\% &51\% & \textit{``What do u think of having a "start cooking" button? Clicking on the recipe name might not be the most intuitive? Thoughts?''}\\
Pro-argument & 17\% & 26\% & 30\% & 23\% & \textit{``It's better UX :) definitely''}\\
Con-argument &18\% & 17\% & 19\% & 18\% & \textit{``But still, I think it is too complicated for now to build it with tabs.''}\\
Decision & 13\% & 7\% & 9\% & 10\% & \textit{``We decided that we will rank our recepies based on frequency of use.''}\\
\bottomrule
\end{tabular}
\label{tab:elements_distribution}
\end{table*}

\subsection{Results}



\begin{figure}
\centering  
\subfigure[Team A]{\includegraphics[width=0.4\textwidth]{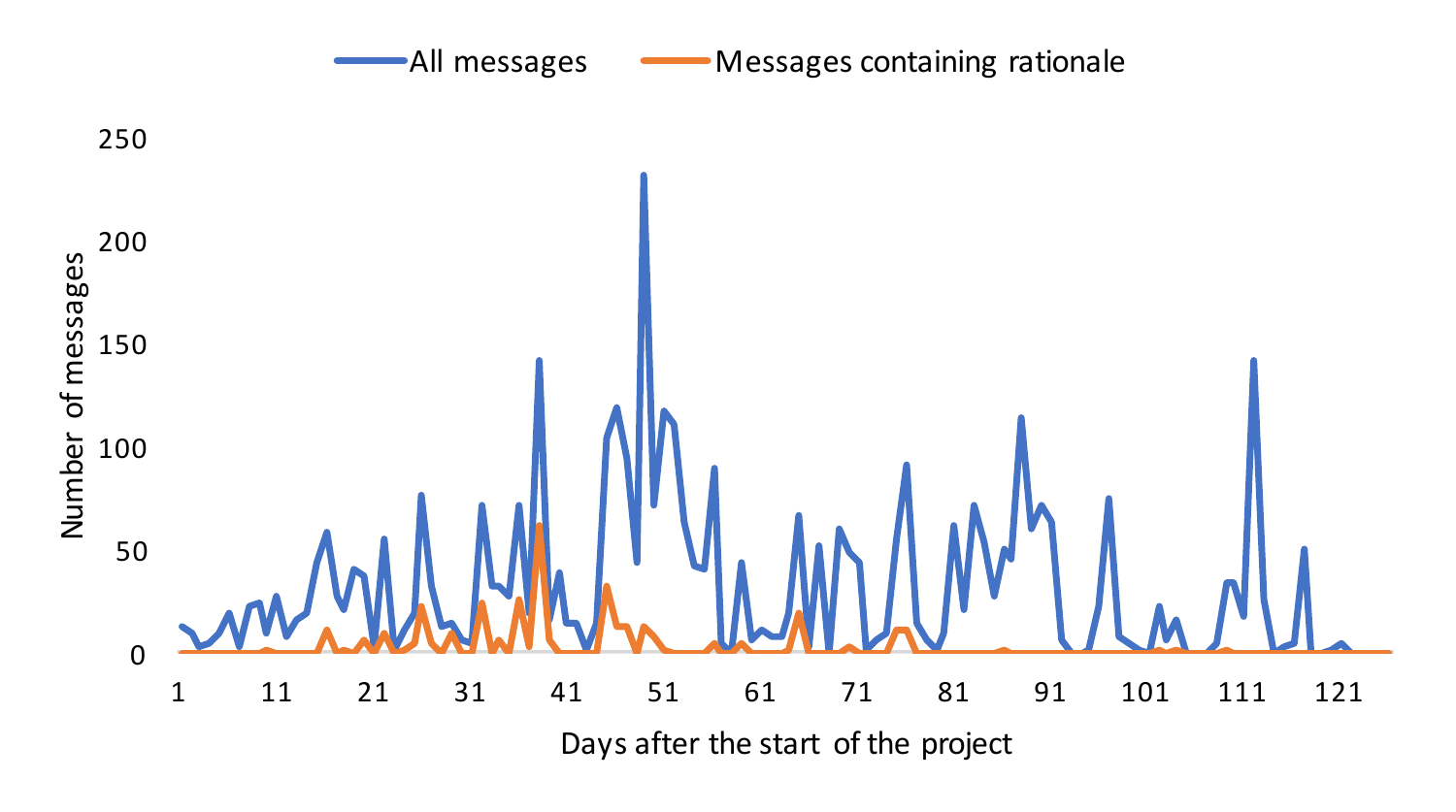}}
\subfigure[Team B]{\includegraphics[width=0.4\textwidth]{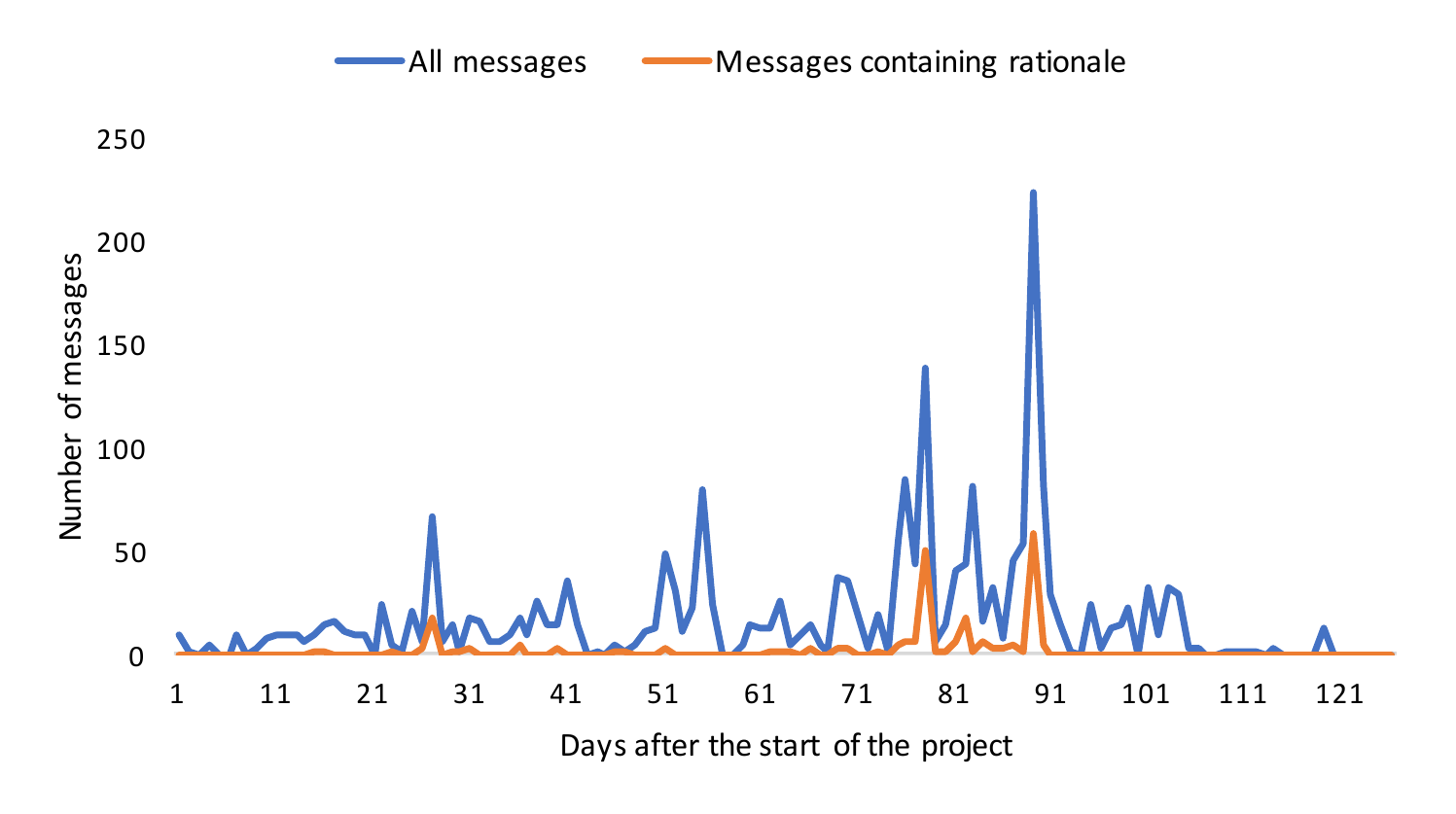}}
\subfigure[Team C]{\includegraphics[width=0.4\textwidth]{{{img/New_Team_C}}}}
\caption{Distribution of all messages and messages containing rationale over the duration of the project.}
\label{fig:time_distribution}
\end{figure}

 Figure~\ref{fig:relevant_msgs} shows the numbers of chat messages identified as containing rationale by the coders. On average 9\% of the team chat messages contain rationale. Although the number of chat messages containing rationale is not high, development teams discuss various elements of rationale in these messages that comprises valuable knowledge about the software system.
 Table~\ref{tab:elements_distribution} shows an overview of the frequency of different rationale elements across messages and examples of coded rationale elements. Overall, we found that proposing alternatives to different issues is predominant in almost 51\% of the messages containing rationale.
 The second most frequent rationale element is issue, present in  24\% of the messages containing rationale.  Pro-arguments were mentioned in 23\% and con-argument in 18\% of the messages containing rationale. These numbers confirm our observation (performed during the manual coding) that team members tend to argue for the alternative they proposed and their reasons to select the alternative (pro-arguments) more frequently than arguing against other alternatives (con-arguments). Lastly, decisions were identified in 10\% of the messages containing rationale. 
 
 To explore how the passage of time and the number of messages influence the amount of rationale found in chat messages, we investigated the distribution of all chat messages and messages containing rationale over the duration of the project. Figure~\ref{fig:time_distribution} visualizes the distribution of all messages and messages containing rationale over the project duration per team. 
 
 We found that different development teams had different rationale distribution in their chat messages. On the one hand, Teams A and B had a significant increase in the number of messages containing rationale at certain time points.
 One possible interpretation of this result is that discussions between developers might be more dense around particular milestones during the project, \eg at the beginning of development sprints or before releasing to the customer. On the other hand, Team C had a steady distribution of chat messages containing rationale that spans the entire project duration. 
 The Spearman's correlation coefficients between the number of days passed since the start of the project and the number of messages containing rationale were 0.19, 0.11 and 0.25 for teams A, B and C respectively indicating a very weak correlation. 
 
 As shown in Figure~\ref{fig:time_distribution}, the increase in the number of chat messages is not always an indicator of an increase in the number of messages containing rationale. However, the Spearman's correlation coefficient between the total number of chat messages and the number of messages containing rationale was 0.5 in all three teams indicating a moderate positive correlation. 
 
 
 

\section{Rationale Completeness}
\label{sec:completeness}

To explore the completeness of rationale in chat messages, we analyzed how many rationale elements were semantically related to other rationale elements also present in the chat messages.



\subsection{Procedure}
Rationale elements identified in different chat messages could be semantically related, as the rationale discussion could span multiple messages. For example, a developer might discuss a specific issue in a single message and other developers propose different alternatives to resolve the issue and argue for and against these alternatives in multiple short messages. Thus, it is important to identify the semantic relationships among these rationale elements in order to explore the completeness of the existing rationale. To achieve this, the same two independent coders who conducted the manual content analysis of the chat messages (detailed in Section~\ref{sec:frequency}) manually inspected the 752 messages containing rationale and identified the semantically related rationale elements contained in these messages. The disagreements were resolved through discussions between the two coders. After identifying the semantically related rationale elements, we answered the following questions:
\begin{itemize}
    \item For each discussed issue, were alternative solutions proposed and was a decision made?
    \item For each selected alternative (made decision), were pro-arguments supporting its selection presented?    
\end{itemize}

\subsection{Results}

Our results show that for 79\% of the issues identified in the chat messages, alternatives were proposed in the chat messages to resolve the issues. In 48\% of these issues, we were able to identify the decisions made, \ie the selected alternatives, in the chat messages.  
However, in only 48\% of the cases in which a decision was made, the pro-arguments in support of the selected alternative were present in the chat messages. Finally, we found that in 21\% of the issues found in the chat messages, neither alternatives were suggested nor a decision has been made. One possible interpretation is that the team members discussed and resolved the open issues through face-to-face communication. For example, while discussing an issue one developer wrote, \textit{“Probably best if we discuss this in the meeting''}. These results support our hypothesis that chat messages are not to be used as the only source for rationale but rather as one of many potential sources. The rationale extracted from chat messages could be integrated with rationale extracted from other sources for a more complete rationale, \eg the decisions made to resolve some of the open issues extracted from chat messages might be available in the meeting minutes of the development team.



\section{Rationale Automatic Extraction}
\label{sec:automatic_extraction}
\begin{figure}
     \centering
 	 \includegraphics[width=0.35\textwidth]{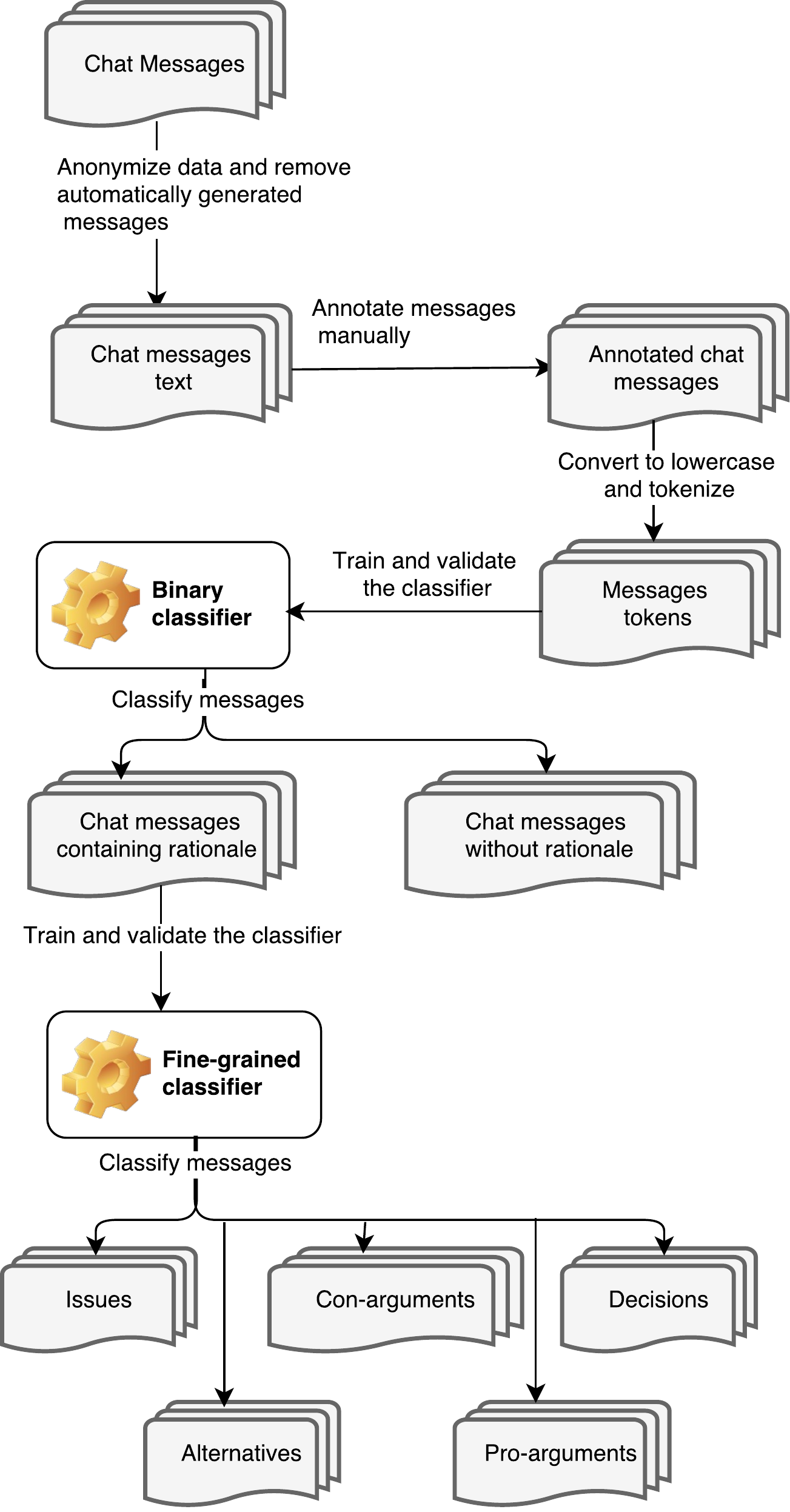}
 	 \caption{Experiment setup.}
 	 \label{fig:process}
 \end{figure}


To explore the  potential  of  automatic  techniques  for rationale  extraction from
chat messages, we performed two experiments. Each experiment classifies rationale, but at different levels of granularity. In the first experiment, we built a \textit{binary classifier} that detects chat messages containing rationale and filters out messages without rationale. In the second experiment, we built a \textit{fine-grained classifier} that classifies the messages containing rationale into the different rationale elements: issues, alternatives, pro-arguments, con-arguments and decisions (defined in Table~\ref{tab:rationale_elements}).

For training and validating our classifiers, we used the manually annotated chat messages described in Section~\ref{sec:frequency}. Figure~\ref{fig:process} shows the experiment setup. We detail each step in the following sections and describe the results. 

\subsection{Binary Classification}

\newcolumntype{g}{>{\columncolor{gray}}l}
\begin{table*}
\centering
\captionof{table}{Binary classification results. } 
\begin{tabular}{llg ggg | ggg }
\toprule
 \textbf{Validation technique}& \textbf{Balancing technique} & \cc& \multicolumn{3}{c}{\textbf{Naive Bayes Multinomial}}  &  \multicolumn{3}{c}{\textbf{SVM}} \\ \cmidrule{4-9}
 & &\cc& \cc Precision & \cc Recall & \cc F1 & 
\cc Precision & \cc Recall & \cc F1 \\ \midrule

\multirow{4}{*}{\textbf{10-fold cross validation}}& &Messages containing rationale & 0.52	&\textbf{0.62}	&\textbf{0.57} &\textbf{0.60}	&0.37	&0.46\\

& & \cc Messages without rationale &\cc \textbf{0.96}	&\cc0.95	&\cc0.95 &\cc0.94	&\cc\textbf{0.98}	&\cc\textbf{0.96}\\ \cmidrule{2-9}

& \multirow{2}{*}{\textbf{Under-sampling}}&  Messages containing rationale&  0.78	&\textbf{0.87}	&\textbf{0.82} &\textbf{0.83}	&0.71	&0.77\\

& & \cc Messages without rationale&\cc  \textbf{0.85}	&\cc0.75	&\cc0.80 &\cc0.75	&\cc\textbf{0.85}	&\cc\textbf{0.80}\\  \cmidrule{2-9}

&\multirow{2}{*}{\textbf{SMOTE + Under-sampling}}& Messages containing rationale& 0.80	&\textbf{0.88}	&0.84 &\textbf{0.85}	&0.85	&\textbf{0.85}\\

& &  \cc Messages without rationale& \cc \textbf{0.87}	& \cc0.78	& \cc0.82 & \cc0.85	& \cc\textbf{0.85}	& \cc\textbf{0.85}\\
\midrule

\multirow{2}{*}{\textbf{Team cross validation}}&&  Messages containing rationale& 0.43	&\textbf{0.61}	&\textbf{0.50} &\textbf{0.48}	&0.28	&0.35\\
&  & \cc Messages without rationale  &\cc \textbf{0.96}	&\cc0.92	&\cc0.94 &\cc0.93	&\cc\textbf{0.97}	&\cc\textbf{0.95} \\

\bottomrule
\end{tabular}
\label{tab:binary_results_message}
\end{table*}


In the first experiment, we explored the potential of the automatic detection of chat messages containing rationale. We applied machine learning techniques to classify the 8,702 chat messages in our dataset into two classes: messages containing rationale and messages without rationale. Considering the large number of chat messages without rationale identified during the manual analysis, the binary classification could be used as a preceding step for a finer-grained classification to filter out message without rationale.

\subsubsection{Experiment Setup}
An important domain in machine learning is document classification, in which documents are assigned to one or more classes. Classifying chat messages into two classes, as in our experiment, is known as binary classification. We compared the performance of two learning algorithms, Naive Bayes Multinomial and Support Vector Machine (SVM) due to their popularity and good performance for text classification~\cite{Frank2006,Sebastiani2002,Chowdhury2015}. 

\paragraph{Preprocessing}
Before training the classifiers, we preprocessed the message text by converting it into tokens and lowercase. Tokenization converts a stream of characters into a sequence of tokens. We used n-gram tokenizer with 1 and 3 as the minimum and maximum length. 
By applying an n-gram tokenizer  we expected patterns of terms appearing together to be indicators of rationale presence in the chat messages, \eg phrases like \textit{“I would suggest''}, \textit{“how about''} could be indicators of proposed alternatives and \textit{“how do we''} could be an indicator of issues.
We chose not to apply stopword removal (\ie removing non-informative, common words) as we expected them to be representative of some rationale elements. For example,  \textit{which} and \textit{how} might be indicators of issues,  \eg \textit{“Which design pattern should we apply?''}, and \textit{but} is commonly used before stating con-arguments against alternatives, \eg \textit{“but it sucks as UX''}. 

\paragraph{Classification level}
We performed our classification on the message level. 
The consideration of the classification on the message level was motivated by two factors. First, previous work on classifying development artifacts~\cite{Rogers2014, Bacchelli2012} found that considering a sentence’s neighbors (context) improved classification performance. Second, chat messages are of short length, which makes classification on the message level feasible.

\paragraph{Data balancing}
Imbalanced datasets are emerging as an important issue in many machine learning applications~\cite{Chawla2004}. Our dataset is imbalanced with an average of 9\% messages containing rationale. Building a classifier from an imbalanced dataset can cause the classifier to be biased towards the majority class, \ie the class with the greater number of instances, while ignoring the minority class~\cite{Han2011}. Two popular techniques for handling class imbalance problem are under-sampling and SMOTE (Synthetic Minority Over-sampling Technique). Under-sampling~\cite{drummond2003c4} uses a subset of the majority class for training the classifier while SMOTE~\cite{Chawla2002} applies oversampling on the minority class by generating synthetic examples. We compared between the application under-sampling and a combination of SMOTE and under-sampling as previous research has proved that classifiers achieve better performance when combining both sampling techniques~\cite{Chawla2002}.

\paragraph{Training and evaluation}
Since we are applying supervised machine learning algorithms, we need to train our classifiers on the manually annotated chat messages (detailed in Section~\ref{sec:frequency}). For training and evaluating the classifiers, we applied 10-fold cross validation and team cross validation. In 10-fold cross validation, 9 folds are used for training the classifier and the remaining fold for validating its performance. The process is repeated 10 times rotating the training and testing folds. The evaluation is computed by calculating the average results among the 10 runs. We also applied team cross validation to better test the generalizability of the results,  since different development teams tend to use different terminologies in their chat messages. In team cross validation, we trained the classifiers on two teams and predicted the classification of the remaining team. We repeated the process three times, \ie a 3-fold cross validation, each time with a different team as a test set and we measured the average results.

We evaluated the classification accuracy using the standard
metrics in machine learning: precision, recall and F-Measure (F1). They are calculated as follows:
\begin{equation}
Precision_i = \frac{TP_i}{TP_i + FP_i} \quad  Recall_i = \frac{TP_i}{TP_i + FN_i}
\end{equation}
Where $TP_i$ is the number of messages that are correctly classified as being of type $i$, $FP_i$ is the number of messages that are incorrectly classified as being of type $i$ and $FN_i$ is the number of messages that are incorrectly classified as not being of type $i$. The F-Measure is the harmonic mean of the precision and recall. 

\subsubsection{Experiment Results}

\begin{table*}
\centering
\captionof{table}{Binary classification results examples. } 
\begin{adjustbox}{max width=\textwidth}
\def\arraystretch{1.5} 
\begin{tabular}{ p{9cm}  p{3cm} p{3cm} }
\toprule
\textbf{Message} & \textbf{Manual classification}&\textbf{Automatic classification}\\
\midrule

\textit{“Well, to be honest you have 3 options. 1) Add new methods to persistency to deal with this new type of data type = redundant code, as it will be deleted. 2) Modify the load all like you mentioned. 3) Just not use persistency for now and, yea, notifications will appear again if the app is closed, but you can test the rest of the functionality.''} &Contain rationale &Contain rationale \\ 

\textit{“You can send it to me on HipChat :)''}&No rationale&No rationale\\


\textit{“found two very very nice tutorials we could use for our graphical stuff (different charts, like pie chart, bar chart, history eg.) and one for scanning qr codes. (Y) i think they look very nice and clean... links''} &No rationale& Contain rationale\\

\textit{“Do we need to support the iPhone4?''}&Contain rationale&No rationale\\

\bottomrule
\end{tabular}
\end{adjustbox}
\label{tab:binary_examples}
\end{table*} 



Table~\ref{tab:binary_results_message} gives an overview of the binary classification results. The numbers in bold represent the corresponding top values.

When applying 10-fold cross validation, both Naive Bayes Multinomial and SVM classifiers performed very well in classifying chat messages without rationale, achieving high values for both precision and recall (above 0.94). However, they performed less well when classifying messages containing rationale; where SVM had a better precision of 0.60 (compared to 0.52 from Naive Bayes Multinomial),  while Naive Bayes Multinomial achieved a much higher recall of 0.62 (compared to 0.37 from SVM). One possible explanation of achieving less accuracy in classifying messages containing rationale is the sparsity of the messages containing rationale in the results of our manual content analysis (Figure~\ref{fig:relevant_msgs}). Table~\ref{tab:binary_examples} shows examples of binary classification results. Upon further inspection, we found that rationale discussions spanning over multiple messages were a common source of error. In these cases, it is important to consider the contextual information in the neighbor messages to identify the rationale contained in the message.  

In the context of extracting rationale from chat messages (rationale was present in only 9\% of the chat messages in our study), we argue that recall is more important than precision. We aim at recovering as  much  rationale  from  chat  messages  as  possible,  with  the compromise  of  falsely predicting  messages  as  containing  rationale. Therefore, we can say that Naive Bayes Multinomial outperformed SVM in classifying messages containing rationale. In summary, both classifiers reported significantly better performance in classifying chat messages without rationale, with the highest achieved precision of 0.96 and recall of 0.98. 
We believe that a binary classifier could be applied as a preceding step for the fine-grained classifier to filter out messages without rationale, as shown in Figure~\ref{fig:process}. 

Applying balancing techniques resulted in a significant increase in the classification performance of messages containing rationale. However, as expected, the performance of classifying the majority class, \ie messages without rationale in our case, decreased. Applying SMOTE in combination with under-sampling achieved higher precision (0.87) and recall (0.88) for classifying messages containing rationale than applying under-sampling alone. 
These results suggest applying  data balancing techniques on the training set to alleviate the classifier bias towards the majority class as a result of imbalanced training data. Consequently, increasing the amount of recovered rationale from chat messages.

When applying team cross validation, the overall classification performance of messages containing rationale decreased. While the Naive Bayes Multinomial classifier achieved an almost equal recall of 0.61 in both validation methods, there were slight decreases in the other accuracy measures. 
The results of classifying messages without rationale are comparable in both validation methods. These findings demonstrate that our results have a high degree of generalizability as we tested the classifier on unseen chat messages of a team different from the teams used for training the classifier and repeated this process three times while rotating the three development teams. 



\subsection{Fine-grained Classification}

\begin{table*}

\centering
\captionof{table}{Fine-grained classification results. } 
\begin{adjustbox}{max width=\textwidth}
\begin{tabular}{lg ggg | ggg | ggg | ggg }
\toprule
&\cc& \multicolumn{6}{c}{\textbf{Binary Relevance}}  &  \multicolumn{6}{c}{\textbf{Label Powerset}} \\ \midrule
 &\cc& \multicolumn{3}{c}{\textbf{Naive Bayes Multinomial}}  &  \multicolumn{3}{c}{\textbf{SVM}} & \multicolumn{3}{c}{\textbf{Naive Bayes Multinomial}}  &  \multicolumn{3}{c}{\textbf{SVM}}\\ \midrule

&\cc& \cc \tiny{Precision} & \cc \tiny{Recall} &\cc \tiny{F1} & 
\cc \tiny{Precision} & \cc \tiny{Recall} &\cc \tiny{F1} &
\cc \tiny{Precision} & \cc \tiny{Recall} &\cc \tiny{F1} &  
\cc \tiny{Precision} & \cc \tiny{Recall} &\cc \tiny{F1} \\ \midrule

\multirow{6}{*}{\textbf{10-fold cross validation}}&Issue & 0.43	&\textbf{0.51}&0.47  &0.51&0.47&0.49   &0.43&\textbf{0.51}&0.47   &\textbf{0.53}&0.47&\textbf{0.50}\\

&\cc Alternative &\cc\textbf{0.67}&\cc0.67&\cc\textbf{0.67}   &\cc0.65&\cc0.58&\cc0.61    &\cc0.66&\cc\textbf{0.68}&\cc\textbf{0.67}     &\cc0.62&\cc\textbf{0.68}&\cc0.65\\
&Pro-argument &0.43	&\textbf{0.53}&\textbf{0.47}      &0.43&0.38&0.40     &\textbf{0.45}&0.37&0.40     &0.41&0.32&0.36 \\

&\cc Con-argument &\cc 0.37&\cc\textbf{0.44}&\cc\textbf{0.40}   &\cc0.37&\cc0.31&\cc0.34   &\cc\textbf{0.39}&\cc0.16&\cc0.23    &\cc0.35&\cc0.23&\cc0.28\\
&Decision & 0.26&\textbf{0.29}&\textbf{0.27}  &\textbf{0.29}&0.21&0.24     &0.23&0.12&0.15     &0.21&0.14&0.17\\

 \midrule

\multirow{6}{*}{\textbf{Team cross validation}}&Issue &0.41&0.45&0.42    &0.50&0.41&0.43     &0.41&\textbf{0.50}&\textbf{0.45} &\textbf{0.51}&0.39&0.43\\

&\cc Alternative &\cc\textbf{0.70}&\cc0.54&\cc0.61    &\cc0.65&\cc0.57&\cc0.60    &\cc0.65&\cc0.69&\cc\textbf{0.66}   &\cc0.60&\cc\textbf{0.70}&\cc0.64\\
&Pro-argument & 0.44&\textbf{0.49}&\textbf{0.45}  &0.42&0.35&0.37  &0.44&0.37&0.38  &\textbf{0.46}&0.29&0.35\\

&\cc Con-argument &\cc 0.28&\cc\textbf{0.32}&\cc0.30   &\cc\textbf{0.38}&\cc0.28&\cc\textbf{0.32}    &\cc0.29&\cc0.12&\cc0.17  &\cc0.35	&\cc0.19&\cc0.24\\
&Decision & 0.22&\textbf{0.32}&\textbf{0.25}  &0.25&0.15&0.16  &\textbf{0.26}&0.14&0.18  &0.22&0.16&0.15\\

\bottomrule
\end{tabular}
\end{adjustbox}
\label{tab:Fine_grained_results}

\end{table*}

\begin{table*}
\centering
\captionof{table}{Fine-grained classification results examples. } 
\begin{adjustbox}{max width=\textwidth}
\def\arraystretch{1.5} 
\begin{tabular}{ p{9cm}  p{3cm}  p{3cm}  }
\toprule
\textbf{Message} & \textbf{Manual classification}&\textbf{Automatic classification}\\
\midrule
\textit{“@User\_3 can we expand on how the app handles lack of internet in the beginning? is there any issue there?''}& Issue& Issue\\
\textit{“Mornin guys. Can someone tell me, what steptype (cooking, measuring, mixing, chopping) preheating the oven is?:/I would make another category - preheating otherwise i would take cooking as the type. What do you guys suggest?''} & Issue, alternative& Issue, alternative\\
\textit{“but it will look like our design and it is easier to implement than our first idea, as we will have one ingredient per line''}&Pro-argument&Pro-argument\\
\textit{“Also it confuses me when the recipe says:"Warm up 50g of sugar and 400ml of cream in a pot."I guess there is one step needed for measuring the stuff and then for cooking them. I have to write in both steps the amount and the ingredients etc. right? That's how i have it in mind.''}& Issue, alternative & Issue, decision\\
\textit{“There should be only one Navigation Controller and the Rest are Views.''}& Decision& Alternative\\

\bottomrule
\end{tabular}
\end{adjustbox}
\label{tab:fine_grained_examples}
\end{table*} 

In the second experiment, we explored the performance of a fine-grained classifier that further classifies the 752  messages containing rationale into five different rationale elements: issue, alternative, pro-argument, con-argument, and decision (defined in Table~\ref{tab:rationale_elements}). The distribution of rationale elements among the messages containing rationale is shown in Table~\ref{tab:elements_distribution}.

\subsubsection{Experiment Setup}
When classifying messages containing rationale into different rationale elements, a chat message might contain more than one element. For example, a developer might propose an alternative and write the pro-argument supporting the alternative in the same message. 
In machine learning, classifying documents into one or more classes that are not mutually exclusive is referred to as multilabel classification. We applied two of the most popular techniques for multilabel classification, the binary relevance (BR) and the label powerset (LP)~\cite{Tsoumakas2007}. In the binary relevance method, a binary classifier is independently trained for each class and the final prediction for a message is determined by aggregating the classification results from all independent classifiers. The main drawback of the binary relevance method is the class independence assumption. 
The label powerset method (LP) takes into account the class correlation by considering each label combination as a single class.

We applied the same preprocessing steps on the message text as used on the binary classification. 
However, we did not apply data balancing techniques because the distribution of rationale elements among the messages containing rationale is more balanced than the data in the binary classification experiment. 

For training and validating the classifiers, we applied both 10-fold cross validation and team cross validation used for the binary classification. In addition, we evaluated the classifiers performance using the same metrics as in the binary classification.

\subsubsection{Experiment Results}

Table~\ref{tab:Fine_grained_results} summarizes the obtained classification results. The numbers in bold represent the corresponding top values.

As expected, the fine-grained classifier -- classifying into five different rationale elements -- performed less well than the binary classifier. When applying 10-fold cross validation, no single classifier works best for all rationale elements. Nevertheless, the results obtained by applying the binary relevance method achieved higher accuracy than when applying the label powerset method in most of the cases. For predicting the different rationale elements, it might be more important to have higher recall values than those of precision to identify as many rationale elements contained in the messages as possible. 
When applying the binary relevance method, Naive Bayes Multinomial outperformed SVM with a higher recall for all rationale elements. We achieved the highest recall for predicting alternatives (0.67). 
Pro-arguments followed with a recall of 0.53, issues with a recall of 0.51 and con-arguments with a recall of 0.44. The accuracy of predicting decisions was the lowest with a recall of 0.29. 

Table~\ref{tab:fine_grained_examples} shows examples of fine-grained classification results. A possible interpretation of the obtained results is the sparseness of some rationale elements in the messages containing rationale (Table~\ref{tab:elements_distribution}). Chat messages are informal short messages and the rationale elements discussed in these messages are unstructured and intertwined. Distinguishing between different elements is a nontrivial and intensive task even for a human judgment. Upon further inspection of the results, we found that a possible interpretation of the poor accuracy in classifying decisions compared to other elements, is that it is not always obvious in the messages whether a decision has been made. And in many cases, the decisions were classified as alternatives by the classifier.  

When applying team cross validation, the overall classification performance of rationale elements decreased.
However, there were slight increases (0.03) in the recall of classifying alternatives and decisions. These results support the generalizability of our results. 

We replicated the above described experiments on the sentence level. 
In both binary and fine-grained classification, the classification on the message level performed significantly better than the classification on the sentence level. Due to the space limitations, sentence level results are not reported.

\section{Discussion}
\label{sec:discussion}
In current agile software development methodologies, the development process is more dynamic and a working software is often given more importance over comprehensive documentation~\cite{beck2001agile}. As a result, informal communication channels, such as chat messages, are seeing wide and rapid adoption by software development teams. These chat messages archive communication between developers and contain valuable information about the rationale behind made decisions. 
This study is a first step towards using chat messages as a source of rationale for the software system.




The results of our study show: (1) chat messages are a valuable source for rationale during software development, (2) chat messages should be used in combination with other development artifacts for capturing a complete rationale, (3) automated filtering of messages without rationale is possible with a high accuracy, and (4) classifying  messages containing rationale into different rationale elements is a promising research direction. In the following we revisit our research questions and discuss possible future work directions.

\textbf{Rationale frequency:} Although a small percentage of chat messages contain rationale (9\%), the manual content analysis results show that these messages contain valuable knowledge about the software system. In chat messages, team members actively engage in discussing issues, proposing alternatives, arguing for and against these alternatives and collaboratively making decisions. 
However, while the informality and unstructured nature of chat messages helps developers reveal their thoughts naturally, this poses a number of challenges for extracting rationale from chat messages. First, rationale could span multiple messages complicating its identification. Second, multiple elements of rationale could be discussed in a single message, and distinguishing between the different elements is a nontrivial task even for a human judgment. 

\textbf{Rationale completeness:} In almost half of the identified issues (48\%), chat messages contained a complete rationale of the alternatives considered, the selected alternative (decision) and the arguments supporting the made decision. However, for the remaining issues, the identified rationale from chat messages was incomplete. A possible explanation is that developers use different communication channels in addition to chat messages. For example, developers might continue some of the chat messages discussions in face-to-face meetings and decisions made to resolve the issues identified in chat messages might be documented in other development artifacts such as meeting minutes. 
This finding emphasizes the importance of linking related rationale elements extracted from different development artifacts for a more complete capturing of rationale. Future research needs to investigate methods and develop tools that systematically extract and aggregate rationale from its identified sources. 

\textbf{Rationale automatic extraction:} Although the manual analysis of chat messages was a useful technique for our research, with the increasing number of chat messages and high volume of messages without rationale, manual analysis becomes infeasible. Therefore, we investigated the use of automated approaches for extracting rationale from chat messages. With the aim of recovering as much rationale from chat messages as possible, our results for detecting rationale in chat messages are encouraging, with a recall ranging from  0.61 to 0.88. 
The filtering of messages without rationale was possible with high precision and recall (both above 0.90). 
However, the results for classifying messages containing rationale into finer-grained rationale elements were less promising. Future work should focus on improving the classification performance by using a more balanced training set with a larger number of manually identified rationale elements. Considering additional features of the messages' text such as using linguistic features and taking into account the message context, \ie neighbor messages, could improve the classification performance.


\textbf{Future work:} Our final goal is to extract rationale from different development artifacts (sources), link related rationale elements and structure rationale in a unified representation. 
Externalizing rationale knowledge can help in achieving a better understanding of the software system and thus supporting future changes. For example, recording explored alternatives could reveal that some proposed changes are inappropriate. Another possible use of rationale is supporting maintenance and design verifications by spotting conceptual and implementation errors~\cite{Dutoit2006}.
Even though a completely automated approach for accurately extracting well-structured rationale is still partially unrealized, our preliminary results of the automatic detection and extraction of rationale from development chat messages are an encouraging step in this direction. Future work could investigate methods for identifying semantic links between messages and structuring rationale from the classified messages into a formal representation, \ie the formalization of rationale from its capture~\cite{Dutoit2006}.

\section{Threats To Validity}
\label{sec:threats_to_validity}

During the manual analysis, the determination if a message contains rationale and what type of rationale elements are present is a subjective decision.
 To mitigate this risk, we created a coding guide with precise definitions and examples for different rationale elements. The guide was used by the coders during the coding task. Furthermore, each message in our dataset was coded by two people and the disagreements were discussed and resolved by the two coders.

Although the list of rationale elements used in our analysis are based on the well-known IBIS model~\cite{KunzundRittel1970}, the list of elements could be incomplete and its descriptions simplified. This threat could lead to the capture of incomplete rationale. However, the used rationale elements are shared among most rationale representation models. Additionally, due to the exploratory nature of our study further extensions and replications are needed.  

We believe that our results have a considerable level of generalizability in regards to other small, agile software projects. The students in the three development teams worked closely with industrial customers and dealt with incomplete and evolving requirements on innovative projects. Previous research found that subject’s experience level might have more effect on the results than the experiment setting (academic or industry) \cite{Salman2015}. In our study, the majority of the student participants described themselves as semi-professional developers and they reported having part-time jobs in the industry. 
Additionally, we conducted team cross validation to test the generalizability of our results regarding the automatic extraction of rationale. 
We encourage further replication of our study to examine if the results reported in this study also hold for different software development settings.

\section{Related Work}
\label{sec:related_work}
We focus the related work discussion in two areas: automatic extraction of rationale and mining developers' communication artifacts. 

\subsection{Automatic Extraction of Rationale}
One method to overcome the rationale capturing  \textit{rationale reconstruction}~\cite{Burge2005}. In this method, the rationale is retrospectively created from development artifacts.
To our best knowledge, this is the first work to explore  the  potential  of automatic  techniques  to  extract rationale   from development team chat messages.  
However, there are several studies on automatic extraction of rationale from other development artifacts. 
Lop\'ez \etal~\cite{Lopez2012} proposed an ontology-driven approach to extract knowledge units relevant for architecture rationale from  plain-text documents. They argue rationale recovery from existing documents can be decomposed into three smaller problems: automatic extraction of rationale from these documents, formalization of the extracted rationale and manipulation of the formalized information for further reuse. Liang \etal~\cite{Liang2012} proposed an algorithm for discovering design rationale from patent documents. They captured rationale in a three layer model consisting of issues, design solutions and artifacts layers. Myers \etal~\cite{Myers1999} designed Rationale Construction Framework (RCF) that recorded designers interactions in a CAD tool and produced a rich history for detailed design. 

Similar to our work, Rogers \etal~\cite{Rogers2012, Rogers2014} applied machine learning techniques to extract rationale from bug reports. In their work, they investigated the use of ontology and linguistic features for training machine learning models to classify sentences containing rationale into decisions, alternatives and argumentation. In their recent work, Rogers~\cite{Rogers2016} proposed a system that uses genetic algorithms to evaluate candidate feature sets for identifying rationale in two types of documents, bug reports and design sessions transcripts. Our work differs from theirs in that we focus on extracting rationale from developers' chat messages. This poses different challenges -- as chat messages are short, informal and less structured than the previously analyzed documents.


\subsection{Mining Developers' Communication Artifacts}
Previous research found that development communication artifacts are a rich source for valuable information about the software system, its history and rationale~\cite{Venolia2006, Wang2012}, as ``developers reveal their thought processes most naturally when communicating with other software developers''~\cite{Seaman1999}. 

Lin \etal~\cite{Lin2016} conducted an exploratory study to understand why developers use Slack, a team messaging platform, and how they benefit from it. Their analysis revealed that most developers use Slack for team-wide purposes including communication and collaboration with other team members.  The findings of their study motivates the work presented in this paper. 

 Previous research has investigated mining Internet Relay Chat (IRC) logs from Open Source Software (OSS) projects. 
Shihab \etal \cite{Shihab2009b,Shihab2009} analyzed the usage of developer IRC meetings channels by project developers and maintainers. 
In their work, they mined IRC meeting logs to investigate the meeting content, meeting participants, their contribution and communication styles.
Chowdhury and Hindle~\cite{Chowdhury2015} implemented machine learning approaches to filter out off-topic discussions in programming IRC channels by exploiting StackOverflow programming discussions and YouTube video comments.  
 Yu \etal~\cite{Yu2011} investigated the use of synchronous (IRC) and asynchronous (mailing list) communication mechanisms in global software development projects. They observed that developers actively use both as complementary communication mechanisms. 
Our work differs from the previous work in that we focus our analysis of chat messages on capturing a specific type of knowledge about the software system, namely, rationale. Furthermore, we analyze chat messages in development settings more similar to commercial than open source software development.

Another stream of research focused on the automated processing of other communication artifacts. Guzman and Bruegge~\cite{Guzman2013} described an approach to summarize collaboration artifacts, such as emails and wikis. Brunet \etal~\cite{Brunet2014}  used machine learning techniques for classifying design discussions in commits, issues and pull requests in open source projects. Similarly, Bacchelli \etal~\cite{Bacchelli2012} and Di Sorbo \etal~\cite{DiSorbo2016} proposed approaches to classify the content of developments' emails into different categories.
Our work builds on these studies and highlights the need for automated techniques to support extraction  and  aggregation of rationale from developers' chat messages.

\section{Conclusion}\label{s:conclusion}
\label{sec:conclusion}
The rationale of the software system is a result of collaboration and negotiation between different stakeholders, as a result developers' discussions in  chat messages are a valuable source for information about rationale. 

In this paper, we report on an exploratory study that investigated rationale frequency, completeness and  the automatic  extraction of rationale  in  two  granularity  levels. We found that  developers discuss  various elements  of  rationale  in these messages, which comprise valuable knowledge about the  software  system. However, due to the high volume of chat messages, automated approaches to extract rationale hidden in these chat messages are needed. The results of our experiments show that applying machine learning techniques can filter out messages without rationale with a precision and recall above 0.90 and that detecting messages containing rationale is possible with a recall ranging from 0.61 to 0.88.
Furthermore, fine-grained classification of the detected rationale is a promising research direction. 
This work is a first step towards a more effective exploitation of rationale in chat messages. 

\section*{Acknowledgments}
We thank Jan Ole Johanßen and Nitesh Narayan for their valuable feedback. This work was partially supported by a PhD scholarship provided by King Saud University for Alkadhi.

\bibliographystyle{abbrv}
\bibliography{references}

\begin{thebibliography}{10}

\bibitem{Bacchelli2012}
A.~Bacchelli, T.~Dal~Sasso, M.~D'Ambros, and M.~Lanza.
\newblock {Content Classification of Development Emails}.
\newblock In {\em Proceedings of the 34th International Conference on Software
  Engineering}, ICSE'12, pages 375--385, 2012.

\bibitem{beck2001agile}
K.~Beck, M.~Beedle, A.~van Bennekum, A.~Cockburn, W.~Cunningham, M.~Fowler,
  J.~Grenning, J.~Highsmith, A.~Hunt, R.~Jeffries, J.~Kern, B.~Marick, R.~C.
  Martin, S.~Mellor, K.~Schwaber, J.~Sutherland, and D.~Thomas.
\newblock {Manifesto for Agile Software Development}, 2001.

\bibitem{Bruegge2009}
B.~Bruegge and A.~H. Dutoit.
\newblock {\em {Object-Oriented Software Engineering Using UML, Patterns, and
  Java}}.
\newblock Prentice Hall Press, 3rd edition, 2009.

\bibitem{Bruegge2012}
B.~Bruegge, S.~Krusche, and M.~Wagner.
\newblock {Teaching Tornado: From Communication Models to Releases}.
\newblock In {\em Proceedings of the 8th Edition of the Educators' Symposium},
  EduSymp'12, pages 5--12, 2012.

\bibitem{Brunet2014}
J.~Brunet, G.~C. Murphy, R.~Terra, J.~Figueiredo, and D.~Serey.
\newblock {Do Developers Discuss Design?}
\newblock In {\em Proceedings of the 11th Working Conference on Mining Software
  Repositories}, MSR'14, pages 340--343, 2014.

\bibitem{Burge2005}
J.~E. Burge.
\newblock {\em {Software Engineering Using RATionale}}.
\newblock {Ph.D. thesis}, Worcester Polytechnic Institute, 2005.

\bibitem{Burge2008}
J.~E. Burge and D.~C. Brown.
\newblock {Software Engineering Using RATionale}.
\newblock {\em Journal of Systems and Software}, 81(3):395--413, 2008.

\bibitem{BurgeJ2008}
J.~E. Burge, J.~M. Carroll, R.~McCall, and I.~Mistr{\'{i}}k.
\newblock {\em {Rationale-Based Software Engineering}}.
\newblock Springer-Verlag, 2008.

\bibitem{Chawla2002}
N.~V. Chawla, K.~W. Bowyer, L.~O. Hall, and W.~P. Kegelmeyer.
\newblock {SMOTE: Synthetic Minority Over-sampling Technique}.
\newblock {\em Journal of Artificial Intelligence Research}, 16(1):321--357,
  2002.

\bibitem{Chawla2004}
N.~V. Chawla, N.~Japkowicz, and A.~Kotcz.
\newblock {Editorial: Special Issue on Learning from Imbalanced Data Sets}.
\newblock {\em ACM SIGKDD Explorations Newsletter}, 6(1):1--6, 2004.

\bibitem{Chowdhury2015}
S.~A. Chowdhury and A.~Hindle.
\newblock {Mining StackOverflow to Filter out Off-topic IRC Discussion}.
\newblock In {\em Proceedings of the 12th International Working Conference on
  Mining Software Repositories}, MSR'15, pages 422--425, 2015.

\bibitem{Cunningham2011a}
H.~Cunningham, D.~Maynard, K.~Bontcheva, V.~Tablan, N.~Aswani, I.~Roberts,
  G.~Gorrell, A.~Funk, A.~Roberts, D.~Damljanovic, T.~Heitz, M.~A. Greenwood,
  H.~Saggion, J.~Petrak, Y.~Li, and W.~Peters.
\newblock {\em {Text Processing with GATE (Version 6)}}.
\newblock 2011.

\bibitem{Cunningham2013}
H.~Cunningham, V.~Tablan, A.~Roberts, and K.~Bontcheva.
\newblock {Getting More Out of Biomedical Documents with GATE's Full Lifecycle
  Open Source Text Analytics}.
\newblock {\em PLoS Comput Biol}, 9(2):e1002854, 2013.

\bibitem{DiSorbo2016}
A.~Di~Sorbo, S.~Panichella, C.~A. Visaggio, M.~Di~Penta, G.~Canfora, and
  H.~Gall.
\newblock {DECA: Development Emails Content Analyzer}.
\newblock In {\em Proceedings of the 38th International Conference on Software
  Engineering Companion}, ICSE'16, pages 641--644, 2016.

\bibitem{drummond2003c4}
C.~Drummond, R.~C. Holte, et~al.
\newblock {C4. 5, Class Imbalance, and Cost Sensitivity: Why Under-sampling
  Beats Over-sampling}.
\newblock In {\em Workshop on learning from imbalanced datasets II}, volume~11,
  2003.

\bibitem{Dutoit2006}
A.~H. Dutoit, R.~McCall, I.~Mistrik, and B.~Paech.
\newblock {\em {Rationale Management in Software Engineering}}.
\newblock Springer-Verlag, 2006.

\bibitem{Fischer1991}
G.~Fischer, A.~C. Lemke, R.~McCall, and A.~I. Morch.
\newblock {Making Argumentation Serve Design}.
\newblock {\em Human-Computer Interaction}, 6(3):393--419, 1991.

\bibitem{Frank2006}
E.~Frank and R.~R. Bouckaert.
\newblock {Naive Bayes for Text Classification with Unbalanced Classes}.
\newblock In {\em Proceedings of the 10th European Conference on Principle and
  Practice of Knowledge Discovery in Databases}, PKDD'06, pages 503--510, 2006.

\bibitem{Guzman2013}
E.~Guzman and B.~Bruegge.
\newblock Towards emotional awareness in software development teams.
\newblock In {\em Proceedings of the 2013 9th Joint Meeting on Foundations of
  Software Engineering}, ESEC/FSE'13, pages 671--674, 2013.

\bibitem{Han2011}
J.~Han, M.~Kamber, and J.~Pei.
\newblock {\em {Data Mining: Concepts and Techniques}}.
\newblock Morgan Kaufmann Publishers Inc., 3rd edition, 2011.

\bibitem{Hesseb2016}
T.-M. Hesse and B.~Paech.
\newblock {Documenting Relations Between Requirements and Design Decisions: A
  Case Study on Design Session Transcripts}.
\newblock In {\em Proceedings of the 22nd International Working Conference on
  Requirements Engineering: Foundation for Software Quality - Volume 9619},
  REFSQ'16, pages 188--204, 2016.

\bibitem{Krusche2014}
S.~Krusche and L.~Alperowitz.
\newblock {Introduction of Continuous Delivery in Multi-customer Project
  Courses}.
\newblock In {\em Companion Proceedings of the 36th International Conference on
  Software Engineering}, ICSE'14, pages 335--343, 2014.

\bibitem{Krusche2014b}
S.~Krusche, L.~Alperowitz, B.~Bruegge, and M.~O. Wagner.
\newblock {Rugby: An Agile Process Model Based on Continuous Delivery}.
\newblock In {\em Proceedings of the 1st International Workshop on Rapid
  Continuous Software Engineering}, RCoSE'14, pages 42--50, 2014.

\bibitem{KunzundRittel1970}
W.~Kunz and H.~Rittel.
\newblock {Issues as Elements of Information Systems}.
\newblock Working Paper 131, Institute of Urban and Regional Development,
  University of California, Berkeley, California, 1970.

\bibitem{Lee1990}
J.~Lee.
\newblock {Artificial Intelligence at MIT Expanding Frontiers}.
\newblock chapter SIBYL: A Qualitative Decision Management System, pages
  104--133. MIT Press, 1990.

\bibitem{Lee1997}
J.~Lee.
\newblock {Design Rationale Systems: Understanding the Issues}.
\newblock {\em IEEE Expert}, 12(3):78--85, 1997.

\bibitem{Liang2012}
Y.~Liang, Y.~Liu, C.~K. Kwong, and W.~B. Lee.
\newblock {Learning the "Whys": Discovering Design Rationale Using Text Mining
  - An Algorithm Perspective}.
\newblock {\em Computer-Aided Design}, 44(10):916--930, 2012.

\bibitem{Lin2016}
B.~Lin, A.~Zagalsky, M.-A. Storey, and A.~Serebrenik.
\newblock {Why Developers Are Slacking Off : Understanding How Software Teams
  Use Slack}.
\newblock In {\em Proceedings of the 19th ACM Conference on Computer Supported
  Cooperative Work and Social Computing Companion}, CSCW'16.

\bibitem{Lopez2012}
C.~L\'{o}pez, V.~Codocedo, H.~Astudillo, and L.~M. Cysneiros.
\newblock {Bridging the Gap Between Software Architecture Rationale Formalisms
  and Actual Architecture Documents: An Ontology-driven Approach}.
\newblock {\em Science of Computer Programming}, 77(1):66--80, 2012.

\bibitem{MacLean1991}
A.~MacLean, R.~M. Young, V.~M.~E. Bellotti, and T.~P. Moran.
\newblock {Questions, Options, and Criteria: Elements of Design Space
  Analysis}.
\newblock {\em Human-Computer Interaction}, 6(3):201--250, Sept. 1991.

\bibitem{mccall1987}
R.~McCall.
\newblock {PHIBIS: Procedurally Hierarchical Issue-based Information Systems}.
\newblock In {\em Proceedings of the International Congress on Planning and
  Design Theory}, volume~44, 1987.

\bibitem{Myers1999}
K.~L. Myers, N.~B. Zumel, and P.~Garcia.
\newblock {Automated Capture of Rationale for the Detailed Design Process}.
\newblock In {\em Proceedings of the Eleventh Conference on Innovative
  Applications of Artificial Intelligence}, AAAI'99, pages 876--883, 1999.

\bibitem{Neuendorf2002}
K.~A. Neuendorf.
\newblock {\em {The Content Analysis Guidebook}}.
\newblock {Sage Publications}, 2002.

\bibitem{Rogers2012}
B.~Rogers, J.~Gung, Y.~Qiao, and J.~E. Burge.
\newblock {Exploring Techniques for Rationale Extraction From Existing
  Documents}.
\newblock In {\em Proceedings of the 34th International Conference on Software
  Engineering}, ICSE'12.

\bibitem{Rogers2016}
B.~Rogers, C.~Justice, T.~Mathur, and J.~E. Burge.
\newblock Generalizability of document features for identifying rationale.
\newblock In {\em Proceedings of the 7th International Conference on Design
  Computing and Cognition}, DCC'16, pages 633--651, 2016.

\bibitem{Rogers2014}
B.~Rogers, Y.~Qiao, J.~Gung, T.~Mathur, and J.~E. Burge.
\newblock {Using Text Mining Techniques to Extract Rationale from Existing
  Documentation}.
\newblock In {\em Proceedings of the 6th International Conference on Design
  Computing and Cognition}, DCC'14, pages 457--474, 2014.

\bibitem{Salman2015}
I.~Salman, A.~T. Misirli, and N.~Juristo.
\newblock {Are Students Representatives of Professionals in Software
  Engineering Experiments?}
\newblock In {\em Proceedings of the 37th International Conference on Software
  Engineering}, ICSE'15.

\bibitem{Seaman1999}
C.~B. Seaman.
\newblock {Qualitative Methods in Empirical Studies of Software Engineering}.
\newblock {\em IEEE Transactions on Software Engineering}, 25(4):557--572,
  1999.

\bibitem{Sebastiani2002}
F.~Sebastiani.
\newblock {Machine Learning in Automated Text Categorization}.
\newblock {\em ACM Computing Surveys}, 34(1):1--47, 2002.

\bibitem{Shihab2009b}
E.~Shihab, Z.~M. Jiang, and A.~E. Hassan.
\newblock {On the Use of Internet Relay Chat (IRC) Meetings by Developers of
  the GNOME GTK+ Project}.
\newblock In {\em Proceedings of the 6th IEEE International Working Conference
  on Mining Software Repositories}, MSR'09, pages 107--110, 2009.

\bibitem{Shihab2009}
E.~Shihab, Z.~M. Jiang, and A.~E. Hassan.
\newblock {Studying The Use of Developer IRC Meetings in Open Source Projects}.
\newblock In {\em Proceedings of the IEEE International Conference on Software
  Maintenance}, ICSM'09, 2009.

\bibitem{Tsoumakas2007}
G.~Tsoumakas and I.~Katakis.
\newblock {Multi-label Classification: An Overview}.
\newblock {\em International Journal of Data Warehousing and Mining}, 3:1--13,
  2007.

\bibitem{Venolia2006}
G.~Venolia.
\newblock {Textual Allusions to Artifacts in Software-Related Repositories}.
\newblock In {\em Proceedings of the 2006 International Workshop on Mining
  Software Repositories}, MSR'06, pages 151--154, 2006.

\bibitem{Wang2012}
H.~Wang, A.~L. Johnson, and R.~H. Bracewell.
\newblock {The Retrieval of Structured Design Rationale for the Re-use of
  Design Knowledge with an Integrated Representation}.
\newblock {\em Advanced Engineering Informatics}, 26(2):251--266, 2012.

\bibitem{Yu2011}
L.~Yu, S.~Ramaswamy, A.~Mishra, and D.~Mishra.
\newblock {Communications in Global Software Development: An Empirical Study
  Using GTK+ OSS Repository}.
\newblock In {\em Proceedings of the 2011th Confederated International
  Conference on On the Move to Meaningful Internet Systems}, OTM'11, pages
  218--227, 2011.

\end{thebibliography}

\end{document}